\documentclass{svjour2}                    

\smartqed  

\usepackage{graphicx}
\usepackage{color}
\usepackage{newlfont}
\usepackage{amssymb}
\usepackage{amsmath}
\usepackage{bbm}
\usepackage[latin1]{inputenc}
\usepackage{cancel}
\usepackage{latexsym}

\newcommand{\ket}[1]{\mbox{$ | #1 \rangle $}}
\newcommand{\bra}[1]{\mbox{$ \langle #1 | $}}
\newcommand{\be}{\begin{eqnarray}}
\newcommand{\ee}{\end{eqnarray}}

\journalname{Foundations of Physics}

\begin{document}

\title{Wave-particle duality: an information-based approach}

\author{R. M. Angelo \and  A. D. Ribeiro}

\institute{R. M. Angelo \at
              Department of Physics, Federal University of Paraná, P.O.Box 19044, 81531-980, Curitiba, PR, Brazil \\
              Tel.: +55-41-33613098\\
              Fax: +55-41-33613418\\
              \email{renato@fisica.ufpr.br}
           \and
           A. D. Ribeiro \at
              Department of Physics, Federal University of Paraná, P.O.Box 19044, 81531-980, Curitiba, PR, Brazil \\
              Tel.: +55-41-33613667\\
              Fax: +55-41-33613418\\
              \email{aribeiro@fisica.ufpr.br}
}

\date{Received: date / Accepted: date}

\maketitle

\begin{abstract}
Recently, Bohr's complementarity principle was assessed in setups involving delayed choices. These works argued in favor of a reformulation of the aforementioned principle so as to account for situations in which a quantum system would simultaneously behave as wave and particle. Here we defend a framework that, supported by well-known experimental results and consistent with the decoherence paradigm, allows us to interpret complementarity in terms of correlations between the system and an {\em informer}. Our proposal offers formal definition and operational interpretation for the dual behavior in terms of both nonlocal resources and the couple work-information. Most importantly, our results provide a generalized information-based trade-off for the wave-particle duality and a causal interpretation for delayed-choice experiments.
\end{abstract}


\section{Introduction} 

Although Bohr's complementarity~\cite{bohr} is widely accepted as a fundamental feature of quantum mechanics, there have been some issues about its precise formulation. While some debate has taken place concerning an eventual dissociation between the complementarity principle and uncertainty relations~\cite{scully91,storey94,englert95}, it seems to be widely accepted by now that there exists an unavoidable connection of complementarity with noncommuting observables~\cite{luis02,busch06} and nonlocality~\cite{oppenheim10,fritz12}. Recently, complementarity was revisited, receiving both theoretical~\cite{terno11,celeri_arxiv12} and experimental~\cite{celeri_pra12,obrien12,tanzilli12} assessments within an interesting framework. Defining the notions of wave and particle in terms of the statistics of clicks in detector settings and using {\em quantum beam splitters}, these works investigated a quantum version of Wheeler's delayed-choice experiment (DCE)~\cite{wheeler78,wheeler84} and concluded that the complementarity principle has to be updated so as to account for a ``morphing behavior''.

\begin{figure}[htb]
\includegraphics[width=\columnwidth]{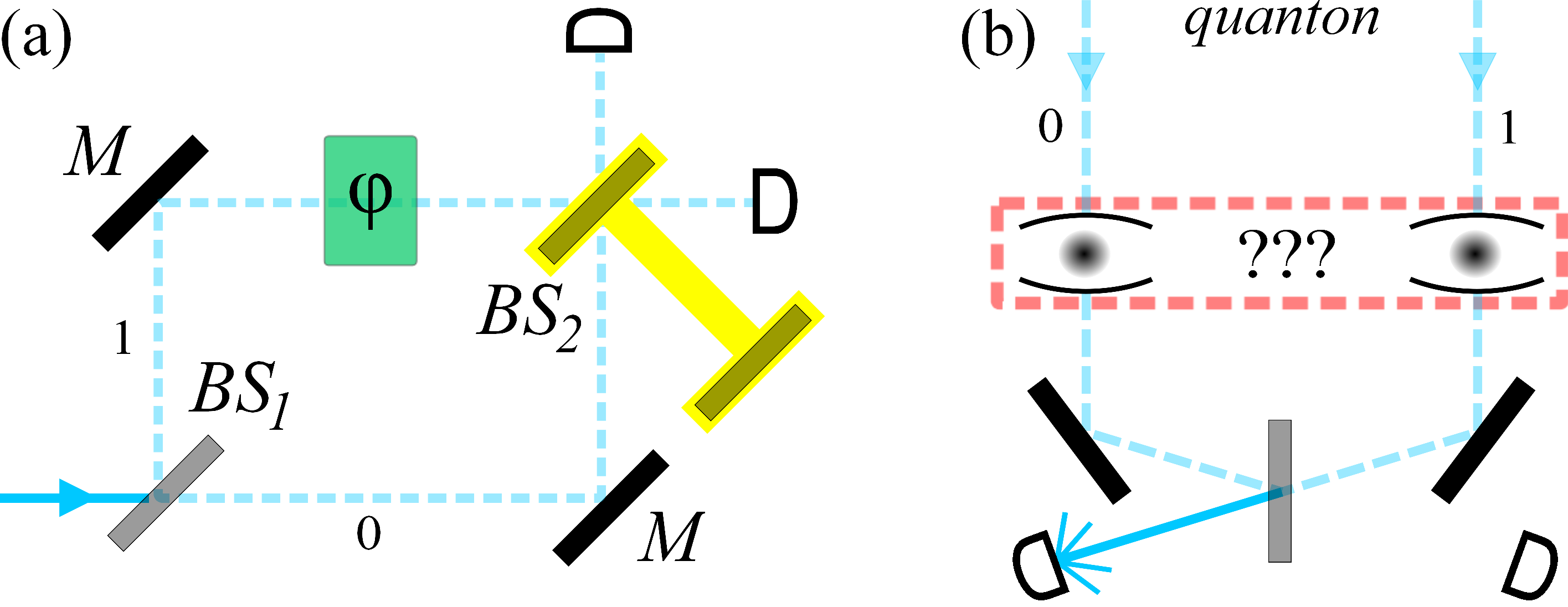}
\caption{(Color online) (a) Mach-Zehnder interferometer composed of two mirrors ($M$), two beam splitters ($BS_{1,2}$),  a tunable phase shifter, and two detectors. The quantum setup is implemented by letting $BS_2$ be in superposition. (b)``Wave detector.'' A quanton is sent towards paths 0 and 1, interacts with two trapped qubits, is redirected by mirrors towards a beam splitter, and finally makes a detector click. When the input state is {\em wavelike}, the qubits get entangled after the click occurs.}
\label{fig1}
\end{figure}

The argument put forward by Refs.~\cite{terno11,celeri_arxiv12,celeri_pra12,obrien12,tanzilli12,wheeler78,wheeler84,girolami12,roch07} can be formulated as follows. A generic quantum system, hereafter called {\em quanton}\footnote{According to J.-M. L\'evy-Leblond~\cite{leblond76}, the term {\em quanton} has been coined by M. Bunge. The utility of this term is in allowing one to refer to a generic quantum system without using words like ``particle'' or ``wave''.}, impinges on a Mach-Zehnder interferometer (MZI) [Fig.~\ref{fig1}(a)] along the path $\ket{0}$. After being split by the beam splitter $BS_1$ into a superposition of distinguishable paths and receiving a relative phase, the quanton ends up in the state
\be 
\ket{p}=\frac{1}{\sqrt{2}}\Big(\ket{0}+i\,e^{i\varphi}\ket{1}\Big). 
\ee 
When the second beam splitter, $BS_2$, is absent, the detectors randomly click with probability $\tfrac{1}{2}$. Since the traveled arm is assumed to be revealed upon a click, $\ket{p}$ is associated with  {\em particlelike} behavior. On the other hand, being present, $BS_2$ recombines the amplitudes and makes the state evolve into
\be 
\ket{w}=\cos\left(\frac{\varphi}{2}\right)\ket{1}-\sin\left(\frac{\varphi}{2}\right)\ket{0}, 
\ee 
up to a global phase. Because the statistics is now sensitive to the phase $\varphi$, implying the occurrence of interference, the quanton is presumed to travel along both arms simultaneously, just like a wave. Then, the state $\ket{w}$ is associated with {\em wavelike} behavior.

By delaying the choice of inserting $BS_2$, Wheeler's proposal aims to defy the assumption according to which some hidden variable would let the quanton know about the state of $BS_2$. Having been ``informed'' about the absence of $BS_2$, the quanton would choose one path, so that it could no longer produce interference even if $BS_2$ were suddenly inserted. Recently, however, such a DCE was realized and interference was observed~\cite{roch07}, in accordance with the statistics associated with $\ket{w}$. Wheeler would have interpreted this result as follows~\cite{wheeler78}: {\em ``Does this result mean that present choice influences past dynamics, in contravention of every formulation of causality? Or does it mean, calculate pedantically and don't ask questions? Neither; the lesson presents itself rather like this, that the past has no existence except as it is recorded in the present.''} 

In the quantum version of the DCE~\cite{terno11,celeri_arxiv12,celeri_pra12,obrien12,tanzilli12,girolami12}, $BS_2$ is prepared in the state $\cos\alpha\,\ket{\text{out}}+\sin\alpha\,\ket{\text{in}}$, which is a superposition of being {\em in} and {\em out} the interferometer [Fig.~\ref{fig1}(a)]. The global state right before the clicks is
\be
\ket{\psi}=\cos\alpha\,\ket{p}\ket{\text{out}}+\sin\alpha\,\ket{w}\ket{\text{in}}.
\label{psi}
\ee
From this state, it is concluded that the complementarity principle must be redefined, since a single setup has been exhibited in which the quanton exhibits a flagrant ``morphing behavior''\footnote{In the Ref.~\cite{terno11}, this term is introduced to name an intermediary behavior, which cannot be exclusively identified either with a wavelike behavior or a particlelike behavior. The notion of a ``morphing behavior'' is induced by a given probability distribution, which is shown to continuously interpolate between a pattern that is fully sensitive to the phase $\varphi$ and a pattern that is fully insensitive to this phase.}. The interpretation underlying this result differs from Wheeler's in essence:{\em ``Behavior is in the eye of the observer''}, for {\em ``particle and wave are not realistic properties but merely reflect how we look at the photon''}~\cite{terno11}. Also, it is in dissonance with a recent analysis claiming that {\em ``each detected photon behaves either as a particle or as a wave, never both, and Bohr's complementarity is fully respected''}~\cite{qureshi13}. 

By basing the diagnostic ``wave or particle'' on the resulting statistics, the conceptual framework developed by the aforementioned works matches, by construction, Bohr's conception of a ``whole unit'', as the quanton behavior gets defined not until the entire setup is arranged. However, the approach faces severe conceptual difficulties, which can be enumerated as follows.

First, those works suggest that the states $\ket{p}$ and $\ket{w}$ are to be associated with the notions of particle and wave, respectively. Indeed, it is under this premise that it is concluded that the entangled state \eqref{psi} reveals a morphing behavior. Now, Schrödinger's equation predicts that, in the region between the two beam splitters, the quanton state is always $\ket{p}$, regardless of the state of $BS_2$. If the quanton route were to be inferred solely from the claimed-particlelike state $\ket{p}$, then one should arrive at the conflicting conclusion that the quanton moves along one path, like a particle, even when $BS_2$ is present. Since we are not willing to deny the validity of Schrödinger's equation, nor wish to supplement the theory with other explanatory elements (e.g., hidden variables), we have to abandon the premise that $\ket{p}$ is to be associated with particlelike behavior. But this immediately invalidates the conclusion of Refs.~\cite{terno11,celeri_arxiv12,celeri_pra12,obrien12,tanzilli12,girolami12} on the meaning of the state \eqref{psi}.

Second, if $\varphi$ is fixed, then it is not possible to verify the phase sensitivity of the statistics, while the query about the route taken by the quanton keeps legitimate. In particular, if $\varphi=\tfrac{\pi}{2}$, then $\ket{p}=\ket{w}$ and the statistics cannot distinguish between these two states. Hence, in this case, these states lose their claimed-different meanings. Furthermore, statistics in the ``interferometer basis'' $\{\ket{0},\ket{1}\}$ are not able to distinguish a superposition $\sqrt{(1-x)}\ket{0}+e^{i\varphi}\sqrt{x}\ket{1}$, which can encode a phase, from a decohered state $(1-x)\ket{0}\bra{0}+x\ket{1}\bra{1}$, which cannot encode any phase. Therefore, such a measurement statistics is not sufficient\footnote{Another example of this limitation is as follows. Consider a double-slit experiment in which $\mathrm{P}_1$ is a plate with only one slit, $\mathrm{P}_2$ is a plate with two slits, and $\mathrm{S}$ a screen where the position of the quanton is recorded. After sending many quantons, one at a time, we register an interference pattern in $\mathrm{S}$. From this statistics, we then infer that, after passing through $\mathrm{P}_1$, the quanton went towards the {\em two} slits in $\mathrm{P}_2$ (it took two paths, as a wave). This is just the standard double-slit experiment. Now, let us put the screen $\mathrm{S}$ very close to $\mathrm{P}_2$. After sending many quantons, we no longer see any interference pattern because the wave packets did not have time to overlap with each other. Then, following the interpretation of Refs. \cite{terno11,celeri_arxiv12,celeri_pra12,obrien12,tanzilli12,girolami12}, one might conclude that, after passing through $\mathrm{P}_1$, the quanton went towards only {\em one} of the slits in $\mathrm{P}_2$ (it took one path, as a particle). But how can it be that the screen, posteriorly located in the space-time structure, has an influence on the quanton when it is passing through $\mathrm{P}_1$? Instead of being an example of retrocausality, this shows that the statistics does not provide a credible inference of the quanton path.} to provide a reliable inference of the quanton route.

Third, as admitted in Ref.~\cite{terno11}, the approach is not able to address the {\em ``tension between the observed interference and the detection of individual photons, one by one, by clicks in the detectors.''} Also, since the detector array is part of the ``whole unit'' that determines the quanton behavior, one would need another detector array (external to this ``whole unit'') to explain the mutually exclusive clicks in the first array. But then we would need a third array to explain why we never see simultaneous clicks in the second, and so on. This logical structure is clearly unsatisfactory. 

There is yet a further difficulty. According to such a statistics-based interpretation, the quanton can behave as a particle even when it is not correlated with any other quantum degree of freedom. In this sense, this approach contrasts with some well-known theoretical developments~\cite{englert96,zurek79,max} which, supported by distinctive experiments~\cite{zeilinger95,haroche96,wineland00}, point out that the particlelike character emerges in the presence of a {\em which-way detector}. In fact, according to the environment-induced decoherence paradigm, it is the entanglement with inaccessible degrees of freedom that prevents interference to occur.

Here, we aim at building a conceptual framework, based on standard concepts of the quantum information field, to address all these issues in a self-consistent way. First, we define the notions of wave and particle in terms of the amount of nonlocality the quanton can activate in a certain setup (Sec.~\ref{operational}). Second, in accordance with this definition, we construct a quantifier that allows one to diagnose the quanton behavior by looking only at its quantum state (Sec.~\ref{measure}). Our measure is shown to have a direct connection with measurement-induced disturbances and the couple information-work. Third, we derive a complementarity relation that establishes a trade-off between the wavelike and particlelike characters, while revealing the role played by the dimension of the quanton Hilbert space (Sec.~\ref{informer}). Our formula, which constitutes a generalization of well-known path-interference duality relations~\cite{qureshi13,englert96,greenberger88}, highlights the role of entanglement in allowing for the emergence of particlelike behavior and shows that morphing behavior already appears in the classical MZI (Sec.~\ref{discussion}). Finally, having offered an entirely forward-causal\footnote{In this work, the notion of {\em causality} (or {\em forward causality}) is to be associated with a time ordering according to which a present state is determined only by past physical instances. In our model, the state of $BS_2$ does not affect the quanton state in an early stage of the experiment, so that any kind of {\em retrocausality} is rejected.} model for duality in terms of standard elements of quantum mechanics, we then conclude that there is no deep concept behind the dual behavior other than noncommutativity and quantum correlations (Secs.~\ref{discussion} and ~\ref{conclusion}).

\section{Operational Definitions\label{operational}} 

In classical physics, a particle is usually described by a vector $\mathbf{r}(t)$, which determines its (definite) position in relation to some origin. A wave, on the other hand, is viewed as a {\em delocalized} disturbance, i.e., something existing simultaneously in many locations. Mathematically, a wave is described by a vector function with space-time dependence, $\mathbf{\Phi}(\mathbf{r},t)$. In the precedent section, these classical notions were employed to discriminate the following two situations: i) when the interferometer was open (i.e., $BS_2$ was absent), no interference was observed, so that the photon should have traveled along only one arm (particlelike behavior); ii) when the interferometer was closed ($BS_2$ present), interference was observed, so that the photon should have traveled along both arms simultaneously (wavelike behavior). However, as mentioned above, this framework presented some inconsistencies and causal issues. In this section, we look for distinct operational definitions that preserve, to some extent, the classical notions of wave and particle, while preserving causality. In particular, we explore the connection between delocalization and Bell nonlocality.

Let us consider the ``wave detector'' depicted in Fig.~\ref{fig1}(b). A quanton traveling along generic paths $\ket{0}$ and $\ket{1}$ is allowed to interact with two trapped qubits, which were prepared in their ground states, $\ket{0}_{0}$ and $\ket{0}_{1}$. The {\em remote} qubits do not interact with each other, but each of them can interact nondestructively with the quanton, i.e., $\ket{0}\ket{0}_0\ket{0}_1\to\ket{0}\ket{1}_0\ket{0}_1$ and $\ket{1}\ket{0}_0\ket{0}_1\to \ket{1}\ket{0}_0\ket{1}_1$. This interaction is manifestly {\em local}, i.e., it only occurs when the quanton crosses the traps. When the quanton enters the wave detector in a superposition of paths, say $\alpha\ket{0}+\beta\ket{1}$, then after the interaction occurs the system state becomes $\alpha\ket{0}\ket{1}_0\ket{0}_1+\beta\ket{1}\ket{0}_0\ket{1}_1$. In this case, tripartite entanglement appears in the system. Note that this will not happen if the quanton is in an incoherent superposition (a mixture) such as $a\ket{0}\bra{0}+(1-a)\ket{1}\bra{1}$ $(a\in[0,1])$. After passing through the traps, the quanton is redirected by two mirrors towards a beam splitter and finally causes a click in one of the detectors. When the quanton enters in superposition, after a click is heard the two-qubit state reduces to $\alpha\ket{1}_0\ket{0}_1\pm i\beta\ket{0}_0\ket{1}_1$, which is a bipartite entangled state. Interestingly, nonlocal correlations have been activated in the two-qubit state only via local interactions. 

From the above, one may conclude that nonlocality has been brought about by the quanton state, i.e., the ``gene'' of the nonlocality was somehow in the incoming state, which was a coherent superposition of paths. This activating mechanism is not present in statistical mixtures because they are not able to leave the qubits entangled. Now, we come to the crux. Two remote qubits can be left nonlocally correlated, via local interactions, only if they are ``touched'' simultaneously by a {\em delocalized} quanton. Here, the analogy with a classical wave seems to be inescapable. Hence, we propose to associate the notion of {\em wave (particle)} with {\em coherent (incoherent)} superpositions of paths. 
 
We now move to a more general situation. Suppose that a quanton enters the wave detector in the state
\be 
\varrho_{\mathcal{Q}}=(1-x)\tfrac{\mathbbm{1}}{2}+x\ket{\psi}\bra{\psi},
\ee 
with $x\in[0,1]$, $\ket{\psi}=\alpha\ket{0}+\beta\ket{1}$, and $|\alpha|^2+|\beta|^2=1$. Assume that the initial state of the system, $\varrho_{\mathcal{Q}}\otimes\ket{0}\bra{0}_0\otimes\ket{0}\bra{0}_1$, evolves to $\tilde{\varrho}$ when the quanton leaves the beam splitter. After a click occurs in the detector $k$ $(k=0,1)$, the state of the two-qubit system conditioned to the click, $\varrho_{\text{\tiny qubits}|k}=\text{Tr}_{\mathcal{Q}}(\Pi_k\tilde{\varrho})/\text{Tr}(\Pi_k\tilde{\varrho})$, reduces to
\be
\varrho_{\text{\tiny qubits}|k}\!&=&\!
\left(\tfrac{1-x}{2}+x|\alpha|^2\right)\!\ket{10}\bra{10}+\left(\tfrac{1-x}{2}+x|\beta|^2\right)\!\ket{01}\bra{01} 
\nonumber \\ &+&(-1)^k \big(i\,x\,\alpha^*\,\beta\ket{01}\bra{10}-i\,x\,\alpha\,\beta^*\ket{10}\bra{01}\big).
\ee
Now, let $\mathcal{N}_l:=\max\{0,\tfrac{1}{4}B_{\text{max}}^2-1\}$ be the maximum degree of violation of a CHSH Bell-inequality achieved via an optimal set of measurements ($B_{\text{max}}$ is defined in Refs.~\cite{hu13,3h95}). For the above state, one shows that $\mathcal{N}_l=4 x^2|\alpha\beta|^2$, which measures the nonlocality activated in the two-qubit system by the quanton. The entanglement produced in the wave detector, $E=2x|\alpha\beta|$, can be computed via concurrence~\cite{yu07}. 

With these results, we set out our physically-motivated definitions. If $x=1$ and $|\alpha|=|\beta|=1/\sqrt{2}$, then nonlocality is maximum and the quanton state will be called {\em strictly wavelike}. On the other hand, when no {\em coherence} is available, i.e., when either $x$ or $\alpha$ or $\beta$ vanishes, nonlocality is minimum, in which case the quanton state will be referred to as {\em strictly particlelike}. Clearly, however, intermediate regimes can exist. In the next section, we provide quantitative statements for these notions.

\section{Waviness Measure \label{measure}} 

We are ready to formalize the notions of particle and wave. To this end, we will employ some well-known tools of the quantum information field. In doing so, we also want to extend the intuition constructed in the previous section. Instead of focusing on the notion of {\em spatial (de)localization}, which was associated with the two ports $\ket{0}$ and $\ket{1}$ of the wave detector, we will adopt the more general concept of {\em (in)definite path}. Let $\{\ket{k}\}$ be the eigenbasis of an observable $\mathcal{K}$ spanning a Hilbert space $\mathcal{H}_{\mathcal{Q}}$ with dimension $d_{\mathcal{Q}}$. For projectors $\Pi_k=\ket{k}\bra{k}$, there holds that $\Pi_k\Pi_{k'}=\Pi_k\delta_{k,k'}$ and $\sum_k\Pi_k=\mathbbm{1}$. When the quanton state is $\ket{k}$, we know with certainty the outcome even before a measurement of $\mathcal{K}$ is performed, so that the quanton is in a {\em definite path}. The notion of definiteness for a mixed state is subtler. Consider that Alice measures the observable $\mathcal{K}$ and then deliveries the quanton to Bob, without telling him the outcome. Aware of the observable measured, Bob is certain that in each run he receives a definite-path state $\rho_k=\Pi_k$ with probability $p_k=\text{Tr}(\Pi_k\rho_{\mathcal{Q}}\Pi_k)$. However, without accessing the information about the outcomes, Bob's prediction for the ensemble is a statistical mixture of definite paths, $\sum_kp_k\rho_k$. By the theorem $S\left(\sum_kp_k\rho_k\right)=H(p_k)+\sum_kp_kS(\rho_k)$~\cite{chuang}, where $S$ and $H$ stand for the von Neumann and Shannon entropies, respectively, we see that $S\left(\sum_kp_k\Pi_k\right)=H(p_k)=-\sum_kp_k\ln p_k$. That is, Bob's lack of knowledge is nothing but subjective ignorance associated with the classical probability distribution $p_k$ secretly prepared by Alice.

These aspects naturally fit our classical intuition, according to which a particle always is in a definite path, even when this path is subjectively ignored. Now, if a quanton is in a definite path state, then its behavior cannot be disturbed by a projective measurement on that path. This motivates us to link the notion of {\em particle} with states that satisfy
\be
\Pi[\rho_{\mathcal{Q}}]=\rho_{\mathcal{Q}},
\label{Pi[rho]}
\ee
where $\Pi[\rho_{\mathcal{Q}}]:=\sum_k \Pi_k\,\rho_{\mathcal{Q}}\,\Pi_k$ describes the unread measurements discussed above (see Refs.~\cite{chuang,luders} for related accounts). If a state $\rho_{\mathcal{Q}}$ is particlelike with respect to $\mathcal{K}$, then it follows from \eqref{Pi[rho]} that $f(\rho_{\mathcal{Q}})=f(\Pi[\rho_{\mathcal{Q}}])$, for a generic $f$. Thus we quantify the extent to which a state $\rho_{\mathcal{Q}}$ is {\em wavelike} upon measurements of $\mathcal{K}$ by 
\be 
I_w(\rho_{\mathcal{Q}}):=\text{Tr}\big[f(\rho_{\mathcal{Q}})-f(\Pi[\rho_{\mathcal{Q}}])\big]. 
\ee 
This gives a generic distance between the state under analysis and a copy subjected to measurements of $\mathcal{K}$. For any differentiable convex function $f$ one shows, by Klein's inequality~\cite{chuang,ana}, that $0\leqslant I_w(\rho_{\mathcal{Q}})\leqslant I_w^{ub}(\rho_{\mathcal{Q}})$, where the upper bound is given by $I_w^{ub}(\rho_{\mathcal{Q}})=\text{Tr}[(\rho_{\mathcal{Q}}-\Pi[\rho_{\mathcal{Q}}])f'(\rho_{\mathcal{Q}})]$. If we assume, in addition, that $f$ is strictly convex, then $I_w(\rho_{\mathcal{Q}})=0$ if and only if $\Pi[\rho_{\mathcal{Q}}]=\rho_{\mathcal{Q}}$. Thus we associate $f$ with measures of information ($I$), generally given by convex functions. We take
\be
\text{Tr}f(\rho_{\mathcal{Q}})=S_{\text{max}}-S(\rho_{\mathcal{Q}})=: I(\rho_{\mathcal{Q}}),
\label{fIS}
\ee
where $S_{\text{max}}$ is the maximum of the entropy $S$ in $\mathcal{H}_{\mathcal{Q}}$. A possible specification for the entropy in Eq.~\eqref{fIS} is the Tsallis entropy~\cite{tsallis88}, $S_q(\rho)=\frac{1-\text{Tr}\rho^q}{q-1}$ $(q>0\in\mathbb{R})$, which is non-negative and strictly concave. It reduces to the von Neumann entropy as $q\to 1$ and recovers the linear entropy $S_2=1-\text{Tr}\rho^2$ as $q=2$. Then, $I_w$ specializes to $I^{\text{\tiny ($q$)}}_w(\rho_{\mathcal{Q}})=S_q(\Pi[\rho_{\mathcal{Q}}])-S_q(\rho_{\mathcal{Q}})$. For $q=2$, a geometric measure derives which is rather convenient for computations~\cite{ana}: 
\be
I_w^{\text{\tiny (2)}}(\rho_{\mathcal{Q}})=\big|\big|\rho_{\mathcal{Q}}-\Pi[\rho_{\mathcal{Q}}]\big|\big|^2,
\label{Iw2}
\ee
where $||\rho||^2:=\text{Tr}(\rho^{\dag}\rho)$ is the square norm in the Hilbert-Schmidt space. On the other hand, the measure resulting for $q=1$ will prove to offer conceptual advantage (see the discussion in the Sec.~\ref{thermo}). Adopting the von Neumann entropy makes the {\em wavelike information} $I^{\text{\tiny ($q$)}}_w$ reduce to
\be
I_w(\rho_{\mathcal{Q}})=S(\Pi[\rho_{\mathcal{Q}}])-S(\rho_{\mathcal{Q}}).
\label{IS}
\ee
The superscript ``(1)'' will be omitted when $q=1$. From Eq.~\eqref{fIS}, we see that the wavelike information is the difference between the information associated with the quanton state and the information available when the quanton is measured.

{\bf \subsection{Thermodynamic Interpretation \label{thermo}}}

Remarkably, $I_w$ possesses a direct thermodynamic interpretation. The key point behind this idea is the link work-information, $W(\rho_{\mathcal{Q}})=k_BTI(\rho_{\mathcal{Q}})$~\cite{zurek03,oppenheim05}, where $k_B$ is the Boltzmann constant and $I(\rho_{\mathcal{Q}})=\ln d_{\mathcal{Q}}-S(\rho_{\mathcal{Q}})$, for a Hilbert space with dimension $d_{\mathcal{Q}}$. The work-information relation gives the amount of work $W$ one can draw from a heat bath of temperature $T$ by the use of the state $\rho_{\mathcal{Q}}$. Along with Eq.~\eqref{fIS}, the work-information relation allows us to express the wavelike information as 
\be
k_BT\,I_w(\rho_{\mathcal{Q}})=W(\rho_{\mathcal{Q}})-W(\Pi[\rho_{\mathcal{Q}}]),
\label{WW}
\ee
whose interpretation is as follows. Suppose that Alice prepares a state $\rho_{\mathcal{Q}}$ and deliveries the quanton to Bob, who can extract an amount $W(\rho_{\mathcal{Q}})$ of work from the heat bath. In a second scenario, the delivered quanton is intercepted by a {\em classical demon}~\cite{zurek03}, who measures $\mathcal{K}$ and forwards the quanton to Bob. Now the extractable work is only $W(\Pi[\rho_{\mathcal{Q}}])$. The wavelike information $I_w$ turns out to be directly related to the difference of work Bob can extract by using $\rho_{\mathcal{Q}}$ and a counterpart secretly accessed by a local Maxwell's demon. Being particlelike, $\rho_{\mathcal{Q}}$ offers no advantage in relation to its measured counterpart.

Before closing this section, it is instructive to compute \eqref{IS} for some states. For $\ket{\psi}=\tfrac{1}{\sqrt{n}}\sum_{k=1}^{n}\ket{k}$ one has that $I_w=\ln n$, which increases with the number of branches in superposition and vanishes for a definite path. For $\ket{w}$ and $\ket{p}$ one gets $I_w(\ket{w})=-x\ln x-y\ln y$, with $x=1-y=\cos^2(\varphi/2)$, and $I_w(\ket{p})=\ln 2$. Since $I_w(\ket{p})\geqslant I_w(\ket{w})$, we may say, in clear dissonance with the interpretation adopted in the Refs.~\cite{terno11,celeri_arxiv12,celeri_pra12,obrien12,tanzilli12}, that $\ket{p}$ is never less wavelike than $\ket{w}$. Finally, we check that our measure is consistent with the diagnostic provided by our wave detector [Fig.~\ref{fig1}(b)]. Using Eq.~\eqref{Iw2}, one shows that 
\be 
I_w^{\text{\tiny (2)}}(\varrho_{\mathcal{Q}})=2x^2|\alpha\beta|^2=2\mathcal{N}_{l}, 
\ee
which highlights our argument that Bell nonlocality can be activated in the wave detector only if the input state has some wavelike character ($I_w^{\text{\tiny (2)}}>0$). Moreover, this relation provides a quantitative connection between waviness and Bell nonlocality.

\vspace{0.75cm}
\section{Complementarity Relation \label{informer}} 

We now discuss a second defining feature of our approach, whereby we ponderate on how waves and particles are produced in nature. To this end, we retrieve the idea according to which coherence is suppressed in the presence of an {\em informer}---an extra degree of freedom which, being correlated with the quanton, can furnish which-path information and makes the quanton behave as a particle. This viewpoint was theoretically identified~\cite{englert96,zurek79,max} and experimentally demonstrated~\cite{zeilinger95,haroche96,wineland00} sometime ago. Here we advance the more radical position which regards that mechanism as {\em the} physical principle underlying the wave-particle duality. 

To elaborate on this view we first note that, whatever the quanton state may be, we can always conceive a purification $\ket{\Psi}\in\mathcal{H}_{\mathcal{Q}}\otimes\mathcal{H}_{\mathcal{I}}$, where $\mathcal{H}_{\mathcal{I}}$ is the (eventually multipartite) Hilbert space of the informer and $\rho_{\mathcal{Q}}=\text{Tr}_{\mathcal{I}}\ket{\Psi}\bra{\Psi}$. It follows that the entanglement $E$ of $\ket{\Psi}$ can be evaluated by $E(\ket{\Psi})=S(\rho_{\mathcal{Q}})$. In search of a complementarity relation, we simply rewrite Eq.~\eqref{IS} as
\be
I_w(\rho_{\mathcal{Q}})+I_p(\rho_{\mathcal{Q}})=I_{\mathcal{H}_{\mathcal{Q}}},
\label{compl}
\ee
where
\be
I_p(\rho_{\mathcal{Q}}):= I(\Pi[\rho_{\mathcal{Q}}])+E(\ket{\Psi})
\label{Ip}
\ee
defines the {\em particlelike information}. In these relations, $I_{\mathcal{H}_{\mathcal{Q}}}:=\ln d_{\mathcal{Q}}$, which gives the maximum information available in the quanton Hilbert space $\mathcal{H}_{\mathcal{Q}}$, and $I(\Pi[\rho_{\mathcal{Q}}])=\ln d_{\mathcal{Q}}-S(\Pi[\rho_{\mathcal{Q}}])$, which is the amount of information accessible via unread measurements of $\mathcal{K}$. 

Equations~\eqref{IS}-\eqref{Ip} give an insightful picture of complementarity. First, we see that the particlelike character is determined by both the measurement-accessible information and the entanglement with an informer. Second, when the entanglement with the informer is maximum ($E=\ln d_{\mathcal{Q}}$), then $I_w=0$ and $I_p=\ln d_{\mathcal{Q}}$, meaning that the quanton behaves as a particle regardless the observable one chooses to measure, i.e., the path of the quanton is definite in all bases. Third, by focusing the analysis on $\rho_{\mathcal{Q}}$, we are able to describe the quanton behavior at any instant of time, in automatic submission to the causal evolution prescribed by Schrödinger's equation. Any change in the quanton behavior is, therefore, constrained to physical interactions modeled in the Hamiltonian of the system, so that any mysterious role of the ``whole unit'' can be physically explained. Fourth, the total amount of information to be distributed between particle- and wave-like behavior is bounded by the dimension $d_{\mathcal{Q}}$ of the quanton Hilbert space. Fifth, our approach can shed some light on the problem of individual clicks in the detectors. We illustrate this point using the following minimalist model of measurement.

Suppose that a quanton is prepared in a wavelike state, $\ket{\mathcal{Q}}=\sum_kc_k\ket{k}$, for which $I_w(\ket{\mathcal{Q}})=H(|c_k|^2)$. Consider an apparatus (informer) prepared in the state $\ket{\uparrow}$. Assume it nondestructively interacts with the quanton via the mapping $\ket{k}\ket{\uparrow}\mapsto\ket{k}\ket{\uparrow_k}$. After the interaction occurs, the joint state becomes $\sum_kc_k\ket{k}\ket{\uparrow_k}$, where $\{\ket{\uparrow_k}\}$ is an orthonormal basis in $\mathcal{H}_{\mathcal{I}}$. Now, it is immediately seen that $I_w(\rho_{\mathcal{Q}})=I_w(\rho_{\mathcal{I}})=0$ in the respective bases $\{\ket{k}\}$ and $\{\ket{\uparrow_k}\}$, since the reduced states are mixtures. This implies that, after the entanglement is established and even before the observer looks at the pointer, both the quanton and the pointer have definite paths. It follows that each one of these systems alone is incapable of producing interference; they individually behave as particles. Notice that, even though one might not conceive the quanton and the pointer as individual systems, as they are entangled, the former is fundamentally inaccessible (this is why we need the latter in the first place) and has to be traced out. When an observer looks at the pointer and gets to know a specific outcome, say $\uparrow_k$, the collapse $\rho_{\mathcal{I}}\to \ket{\uparrow_k}$ occurs. Then, after reading the pointer, this observer concludes that both the pointer and the quanton behave as particle, as $I_w(\ket{\uparrow_k})=I_w(\ket{k})=0$. Interestingly, we see that the act of looking at the pointer does not change the definiteness of paths, but only reveals an ignored path. In this sense, the collapse is nothing but information updating. It is also worth noticing that while the wavelike information of the quanton changes during the measurement process, the pointer preserves its particlelike identity. Now, let us consider an instance in which two detectors (pointers) are involved. Tracing the quanton out yields a reduced state in the form $p_{\text{\tiny RC}}\ket{\text{R}}\bra{\text{R}}\otimes\ket{\text{C}}\bra{\text{C}}+p_{\text{\tiny CR}}\ket{\text{C}}\bra{\text{C}}\otimes\ket{\text{R}}\bra{\text{R}}$ (R=ready, C=click), with $p_{\text{\tiny RC}}+p_{\text{\tiny CR}}=1$. Such a classically correlated mixed state admits an interpretation according to which {\em either} one {\em or} the other detector clicks alone. This simple model\footnote{After the completion of this work, we became aware of Ref.~\cite{hobson15}, which proposes a solution to the measurement problem using related arguments.}, which can be straightforwardly extended to any system involving a quanton and a set of detectors, shows that, due to entanglement, there can be no two simultaneous clicks in a set of detectors\footnote{In this model, environment-induced decoherence is not required to explain why the pointer manifests in a particlelike manner, although it would be useful to account for the irreversibility of the measurement process.}.

The discussion on the genesis of the dual behavior can now be concluded. Consider an entangled state $\sum_kc_k\ket{k}\ket{\mathcal{I}_k}$, where $\ket{\mathcal{I}_k}\in\mathcal{H}_{\mathcal{I}}$. Since $I_w(\rho_{\mathcal{Q}})=0$ if and only if $\rho_{\mathcal{Q}}=\sum_kp_k\Pi_k$, strictly particlelike behavior will occur only if $\langle\mathcal{I}_{k'}|\mathcal{I}_k\rangle=\delta_{k',k}$. In this case, the information of every quanton state $\ket{k}$ is exclusively encoded in an informer state $\ket{\mathcal{I}_k}$. From the viewpoint of the informer, therefore, the situation is such that the quanton is always in a definite path\footnote{This point can be easily illustrated for the superposition $\ket{\phi}=\alpha\ket{x_{\mathcal{Q}}}\ket{x_{\mathcal{I}}}+\beta\ket{x_{\mathcal{Q}}+\delta}\ket{x_{\mathcal{I}}+\delta}$, where $\ket{x_{\mathcal{Q},\mathcal{I}}}$ denotes a sharp Gaussian state centered at $x_{\mathcal{Q},\mathcal{I}}$. In terms of center-of-mass and relative coordinates this state reads $\ket{\phi}=\left( \alpha\ket{x_{cm}}+\beta\ket{x_{cm}+\delta}\right)\ket{x_{\mathcal{Q}}-x_{\mathcal{I}}}$, where $x_{cm}=(m_{\mathcal{Q}}x_{\mathcal{Q}}+m_{\mathcal{I}}x_{\mathcal{I}})/(m_{\mathcal{Q}}+m_{\mathcal{I}})$. Clearly, as far as the relative physics is concerned, the quanton is in a definite path. See Refs.~\cite{angelo11,angelo12} for related discussions.}. Nevertheless, there  is no further informer witnessing the path of the informer itself, so that there is still a lack of information for the joint state. This explains how entanglement makes the quanton path definite while the joint state remains a superposition of paths. On the other hand, if $\rho_{\mathcal{Q}}=\sum_kp_k\Pi_k$, one can build a purification $\sum_kc_k\ket{k}\ket{\uparrow_k}$, with $p_k=|c_k|^2$, showing that particlelike behavior can always be thought of as emerging due to entanglement with an informer. If entanglement is absent, genuine particlelike behavior appears only when the quanton is in an eigenstate of $\mathcal{K}$. However, as seen in our measurement model, as far as information is concerned, a collapsed state $\ket{k}\ket{\uparrow_k}$ is physically equivalent to a ``non-collapsed'' entangled state $\sum_kc_k\ket{k}\ket{\uparrow_k}$, with $\ket{\uparrow_k}$ being a pointer state. Indeed, in both cases the wavelike information of the quanton and the informer is zero. Once again, entanglement proves to be crucial. 

The wavelike behavior, by its turn, is favored in the following situation. Let $\gamma_{\mathcal{Q}}+\gamma_{\mathcal{I}}=\Gamma$ be a generic conservation law obeyed by the Hamiltonian. Particlelike states, such as $\ket{\gamma_{\mathcal{Q}}}\ket{\gamma_{\mathcal{I}}}$, may dynamically evolve into $\alpha\ket{\gamma_{\mathcal{Q}}}\ket{\gamma_{\mathcal{I}}}+\beta\ket{\gamma_{\mathcal{Q}}+\delta\gamma}\ket{\gamma_{\mathcal{I}}-\delta\gamma}$, so as to satisfy the conservation law. In this case, the wavelike information for the quanton, computed via Eq.~\eqref{Iw2}, reads $I^{\text{\tiny (2)}}_w(\rho_{\mathcal{Q}})=2|\alpha\beta|^2|\langle \gamma_{\mathcal{I}}-\delta\gamma|\gamma_{\mathcal{I}}\rangle|^2$. It is clear that strictly wavelike behavior will emerge only if $|\alpha|=|\beta|=1/\sqrt{2}$ (equally probable paths) and $|\langle \gamma_{\mathcal{I}}-\delta\gamma|\gamma_{\mathcal{I}}\rangle|\approx 1$. The latter condition is fulfilled whenever $\delta\gamma$ is much smaller than the width of $\ket{\gamma_{\mathcal{I}}}$, a situation that occurs when the interaction is not able to significantly disturb the informer state. This is precisely what happens, e.g., when a quanton passes through either a 50:50 beam splitter or a double-slit system: Momentum-conserving interaction of a quanton with an undisturbable object makes the former behave like a wave. This scenario prompts us to propose a primitive statement for the underlying physics of duality: {\em Genuine particlelike behavior emerges only when a quanton is entangled with an informer.} 

\section{Discussion\label{discussion}}

As far as DCEs are concerned, within our framework the logic of ``delaying a choice'' represents no issue at all. Immediately after passing through $BS_1$, the quanton can never behave as particle because there is no informer in the system. The posterior introduction of $BS_2$, eventually in a far distant future, cannot change the present state of the quanton, whose physics is defined only by the interaction with $BS_1$. The presence of $BS_2$ will demand, on the theoretical side, the inclusion of a potential in the Hamiltonian. This done, one will be able to predict that, upon the interaction with $BS_2$, the quanton behavior will causally change. While no object is included in the system, nothing changes. 

In the quantum MZI [Fig.~\ref{fig1}(a)], after an interaction occurs between the quanton and $BS_2$, the situation is as follows. Tracing $BS_2$ out from state \eqref{psi}, one shows that $I_p^{\text{\tiny (2)}}(\rho_{\mathcal{Q}})=\tfrac{1}{2}(1-\cos^4\alpha)\cos^2\varphi$ and $E^{\text{\tiny (2)}}=\tfrac{1}{4}\sin^2(2\alpha)\cos^2\varphi$, where the entanglement is now measured via linear entropy. If $BS_2$ is in superposition $(\alpha\neq 0,\tfrac{\pi}{2})$, then $I_p^{\text{\tiny (2)}}(\rho_{\mathcal{Q}})$ never reaches the maximum value $\tfrac{1}{2}$, which means that the quanton cannot behave as a particle until reaching the detectors. Morphing behavior will generally occur after $BS_2$, but this is also the case in the classical MZI $(\alpha=\tfrac{\pi}{2})$, for which the quanton behavior, as predicted by $I_{w}^{\text{\tiny (2)}}=\tfrac{1}{2}\sin^2\varphi$ and $I_{p}^{\text{\tiny (2)}}=\tfrac{1}{2}\cos^2\varphi$, can be directly adjusted by the phase $\varphi$.

It is worth noticing that the measure $I_{w,p}$ is a {\em relational} quantifier, i.e., it depends on the {\em observable} one uses to probe it. For instance, an eigenstate of $\sigma_z$, say $\ket{0}$, will be diagnosed as particlelike upon measurements of $\sigma_z$ and wavelike upon measurements of $\sigma_x$. This is so because, in general, a quantum state does not present the same degree of coherence in all bases. Indeed, this conclusion is consistent with the dual notion formulated in our wave detector, as the capability of the state $\ket{0}$ to activate nonlocality crucially depends on how we orientate the Stern-Gerlach (SG) in order to previously separate the paths. The eigenstate $\ket{0}$ will not split in a superposition of paths after passing a SG aligned with the $z$ axis, so that it cannot activate nonlocality in the wave detector. Conversely, it will split and activate nonlocality when SG is aligned in the $x$ direction. The fact that the wavelike information is relational derives from the incompatibility of observables. This can be formally expressed by the easily-provable relation\footnote{As pointed out by an anonymous referee, Eq. \eqref{JK} can be viewed as an alternative expression of L\"uders's theorem~\cite{luders}. We thank the referee for this remark.}
\be
\mathcal{J}-\Pi[\mathcal{J}]=\sum_k\,[\mathcal{J},\Pi_k]\,\Pi_k,
\label{JK}
\ee
which shows that an arbitrary operator $\mathcal{J}$ will not be disturbed upon measurements of $\mathcal{K}=\sum_kk\Pi_k$ if and only if it commutes with the latter. This remark illustrates how our approach links to the issue of the complementarity of noncommuting observables. More insightful formulations can be constructed in specific cases. Consider, for instance, the state $\ket{\psi}=a\ket{0}+b\,e^{i\theta}\ket{1}$, with $a^2+b^2=1$, as written in the $\sigma_z$ basis. From the results of Sec.~\ref{thermo}, we get $I_w^{\text{\tiny (2)}}[\sigma_z]=2a^2b^2$. Here, the notation explicitly indicates the measured observable. Considering measurements of $\sigma_{x,y}$, the wavelike information results $I_w^{\text{\tiny (2)}}[\sigma_{x,y}]=\frac{1}{2}-2a^2b^2\cos^2\theta$, which is clearly sensitive to the phase $\theta$. From these relations, one immediately obtains $\sum_s I_w^{\text{\tiny (2)}}[\sigma_s]=1-2a^2b^2\cos(2\theta)$ ($s=x,y,z$), which is lower bounded as
\be
I_w^{\text{\tiny (2)}}[\sigma_x]+I_w^{\text{\tiny (2)}}[\sigma_y]+I_w^{\text{\tiny (2)}}[\sigma_z]\geqslant \frac{1}{2}.
\ee
Interestingly, this inequality constitutes a clear statement of the complementarity of noncommuting observables for the state $\ket{\psi}$. It shows that the waviness of a pure state cannot be removed in all measurement bases.

In a different vein, one may wonder whether information is also {\em relative}, i.e., if for a given reference observable, distinct {\em observers} would access different values for $I_{w,p}$, while predicting the same physics. This problem was recently investigated in the context of quantum references frames~\cite{angelo12}. Considering a double-slit experiment inside a very light lab whose movement would constitute an informer for an external observer, and adopting a relative formulation for quantum states, the authors discussed how the same physical fact manifests in the viewpoint of each observer. Their approach and conclusions are in full agreement with the primitive elements adopted here for duality. 

Finally, we point out that like Englert's approach~\cite{englert96}, ours admits both the fundamental role of an informer and the occurrence of morphing behavior (here signalized when $I_{w}>0$ and $I_{p}>0$ simultaneously). As advantages, our formulation relies on an equality [Eq.~\eqref{compl}], directly applies to Hilbert spaces of arbitrary dimensions, and is based on an information-theoretic measure. This result extends previously reported duality relations~\cite{qureshi13,englert96,greenberger88}.

\section{Conclusion \label{conclusion}} 

In summary, employing standard tools of the quantum information theory, we propose a conceptual framework for duality that naturally explains all the recent experimental results and avoids conceptual puzzles. We abdicate Bohr's notion of a ``whole unit'' in favor of a model that links waviness with the coherence of the quanton state. This approach proved to be rather fruitful: Besides offering an interesting connection between the wavelike behavior and two fundamental physical concepts, namely, Bell nonlocality and thermodynamic work, it allows us to consistently account for the problem of individual clicks in detector settings. Importantly, it also yields an information-based relation that generalizes some well-known duality relations. Our approach relies {\em only} on primitive elements of the standard quantum theory, namely, deterministic evolution (Schrödinger's equation), physical causation (interactions), correlations (the role of the informer), and partial trace (for the diagnostic of subsystems). It follows that, in contrast to recent works claiming further sophistication for the complementarity principle, our framework corroborates the perspective according to which Bohr's complementarity is nothing but a consequence of noncommutativity and quantum correlations.

\begin{acknowledgements}
This work was supported by the project National Institute for Science and Technology of Quantum
Information (INCT-IQ; CNPq/Brazil). We thank L. C. Celeri and P. Milman for helpful discussions. 
\end{acknowledgements}


\end{document}